\documentstyle[12pt]{article}
\begin {document}

\title{NONRELATIVISTIC MODEL FOR $b\overline{b}$ QUARKONIA \footnote{ talk at
the Workshop on heavy quarks held in Bad Honnef 14 -- 17 December 1994}}

\author{
L. MOTYKA and K. ZALEWSKI\\
Institute of Physics, Jagellonian University, \\
ul. Reymonta 4, Krak\'ow, 30 059 Poland }

\maketitle

\begin{abstract}
{Experimental data for $b\overline{b}$ quarkonia have been compared
with the predictions of a variety of nonrelativistic quark models. It is found
that a potential $a\sqrt{r} - b/r +$ const gives good agreement, while many
others do not. Some implications of this observation are discussed.}
\end{abstract}

\section{Introduction}
According to a recent review\cite{BES} the masses of the $b \overline{b}$
quarkonia below twice the mass of the $B$ meson are successfully predicted by a
variety of potential models. The average error of the predictions is typically
a few MeV --- $2.3$ MeV in the best case\cite{FUL}. Some of these models
include relativistic corrections, but if we replace the masses of the $\chi$
multiplets by their centres of gravity, the relativistic corrections are not
crucial for the success. This result is puzzling. From fine splittings, which
are a purely relativistic effect, or from the differences between typical
kinetic energies evaluated relativistically and nonrelativistically, one finds
that the relativistic corrections should be of some tens of MeV. The standard
explanation is that what is called the nonrelativistic Hamiltonian, is in fact
an effective Hamiltonian with the parameters modified by relativistic
corrections. This rises the question: is it possible to reproduce the data
within errors using such an effective Hamiltonian? Since experimental errors on
the masses are of the order of $0.3$ MeV, even the "successful" models
correspond to very low confidence levels as evaluated from the $\chi^2$ test.
Thus in order to answer the question one should do better. Moreover, there are
also other data, e.g. leptonic widths and electric dipole transition
probabilities, which could and should be fitted. Quark - antiquark potentials
derived from studies of quarkonia are used in a variety of applications. Let us
mention as examples applications to heavy light systems\cite{ISG}, to
$b\overline{c}$ mesons\cite{FLS}, and to $t\overline{t}$ production\cite{JEZ}.
 Therefore, improving on existing potentials seems useful. On the
other hand, as is well known, two body relativistic quantum mechanics does not
exist yet. If no nonrelativistic potential can reproduce the data, this
may be a valuable hint on how to construct a relativistic theory.

\section{The model} There is no standard nonrelativistic quarkonium model. The
Schr\"odinger equation is

\begin{equation}
- \frac{1}{m_b}\vec{\nabla}^2 \psi + V(r) \psi = E \psi,
\end{equation}
but the quark mass $m_b$ and the potential $V(r)$ vary from paper to paper.
Fortunately, most proposals for the potential either are of the form

\begin{equation}
V(r) = -a r^{-\alpha} + b r^\beta + Ct,
\end{equation}
where $a,b,\alpha, \beta, Ct$ are nonnegative constants, or are numerically
very well approximated by such formulae. Let us mention as examples the Cornell
potential\cite{EIC} with $\alpha = \beta = 1$, the potential of Lichtenberg
and collaborators\cite{LIC} with $\alpha = \beta = 0.75$, the potential of Song
and Liu\cite{SOL} with $\alpha = \beta = 0.5$, the logarithmic potential of
Quigg and Rosner\cite{QUR} corresponding to $\alpha = \beta \rightarrow 0$ and
the potential of Martin\cite{MAR} corresponding to $\alpha = 0,\; \beta = 0.1$.
Other potentials, which have sometimes quite different analytic forms, become
very close to potentials of this type, when their parameters are adjusted to
fit the data. We have checked  that in particular for the very successful
Indiana potential\cite{FOG} and for the famous Richardson potential\cite{RIC}.
 We found that it is enough to impose on all these potentials the
condition that they fit the energy spectrum as well as possible within the
freedom described below. This is related to the suggestion of Eichten
and Quigg, who concluded from their inverse scattering analysis\cite{QUR1}
that the first few $L=0$ energy levels and the corresponding leptonic widths
(not used by us, but we use instead the energies of the $L=1$ levels) to a
large extent determine the potential in the region relevant for the quarkonium
calculations. Note, however, that we have no proof that a completely different
potential would not fit the data as well.

In order to make our search for a satisfactory potential more effective, we
introduced a scaled variable $\rho = \lambda r$. In terms of this variable the
Schr\"odinger equation (1) with potential (2) can be reduced to the form

\begin{equation}
-\vec{\nabla}^2 \psi(\rho) + {\cal V}(\rho) \psi(\rho) = {\cal E} \psi(\rho),
\end{equation}
where

\begin{eqnarray}
\lambda & = & \left(\frac{b}{a}\right)^{\frac{1}{\alpha + \beta}},\\
{\cal V}(\rho) & = & C( \rho^\beta - \rho^{-\alpha}),\\
{\cal E} & = & \frac{m_b}{\lambda^2}(E - Ct),\\
C & = & m_b a \lambda^{\alpha - 2}.
\end{eqnarray}
Thus, for given values of the parameters $\alpha, \beta$ the model has four
free parameters, which can be chosen as $m_b$, $\lambda$, $Ct$, and $C$. Our
idea is to choose a set of observables, which depend only on the parameter $C$.
Parameter $C$ is chosen so as to minimize $\chi^2$ for the observables chosen.
Then other observables can be used to determine the remaining parameters of the
model, but this does not influence the quality of the fit. Using this procedure
for the potentials proposed in the literature we got significant improvements
of the fits, but none was quite satisfactory.

\section{Observables}
As observables we have chosen

\begin{eqnarray}
b_1 & = & \frac{M(2S) - M(1S)}{M(3S) - M(1S)}  =  0.6290 \pm 0.0005,\\
b_2 & = & \frac{M(3S) - M(2P)}{M(2S) - M(1P)}  =  0.774 \pm 0.006,\\
b_3 & = & \frac{M(2S) - M(1P)}{M(2S) - M(1S)}  =  0.219 \pm 0.001,\\
b_4 & = & \frac{|\psi_{2S}(\vec{0})|^2}{|\psi_{1S}(\vec{0})|^2}  =  0.492 \pm
0.111,\\
b_5 & = & \frac{|\psi_{3S}(\vec{0})|^2}{|\psi_{1S}(\vec{0})|^2}  =  0.433 \pm
0.071,\\
b_6 & = & |\psi_{1S}(\vec{0})|^{\frac{2}{3}}\langle 1P| r | 2S \rangle  =
2.29 \pm 0.16, \\
b_7 & = & |\psi_{1S}(\vec{0})|^{\frac{2}{3}}\langle 2P| r | 3S \rangle  =
1.59 \pm 0.15, \\
b_8 & = &  \frac{\langle 1S| r | 2P \rangle}{\langle 2S| r | 2P \rangle}  =
0.110 \pm 0.009.
\end{eqnarray}

The observables $b_1,\ldots,b_8$ depend on the parameter $C$, but not on the
other three parameters. Thus the $\chi^2$ distribution corresponds to seven
degrees of freedom. All the numerical values are calculated from the data given
in the 1994 Particle Data Group Tables\cite{PDG}. There are some problems with
the observables $b_6$ and $b_7$, however. The formulae connecting the measured
leptonic widths to the calculated $|\psi_{nS}(\vec{0})|^2$ contain a radiative
correction factor. This is $1 - \frac{16 \alpha_s}{3 \pi}$ and we put it equal
0.7. Since neither the exact value of $\alpha_s$, nor the effect of the higher
order terms is known, this number is uncertain by perhaps some 20\%.. This
uncertainty very probably cancels in $b_4,\; b_5$ and $b_8$ , but for $b_6$ and
$b_7$ it introduces a common factor, which deviates from unity by some seven
per cent. This uncertainty has not been included in the errors. Note also that
the simple dipole formulae are much less reliable for quarkonia than for atoms.

Once the parameter $C$ is chosen so as to reproduce as well as possible the
observables $b_1,\ldots,b_8$, the parameters $\lambda$, $m_b$ and $Ct$ can be
calculated from the experimental value the leptonic width of the
$\Upsilon(1S)$ state, which yields $|\psi_{1S}(\vec{0})|^2$, from the mass
difference $M(3S) - M(1S)$ and from the mass $M(1S)$. This calculation has no
bearing on the quality of the model, except that perhaps values of $m_b$ too
far from $5$ GeV would make the model unplausible.

\section{Results and conclusions}
After some exploration in the $\alpha, \beta$ plane, we have found a positive
answer to our question. For $\alpha = 1,\; \beta = 0.5$ we find for $C =
0.9315$ the value $\chi^2 = 6.5$, which is very satisfactory for the seven
degrees of freedom. This solution corresponds to the potential in eq. (1) given
by

\begin{equation}
V(r) = 0.70585\left(\sqrt{r} - \frac{0.46122}{r}\right),
\end{equation}
where $V(r)$ and $r$ are in GeV. The corresponding quark mass $m_b = 4.79333$
GeV is quite reasonable. For $r \rightarrow 0$ our potential has the $r^{-1}$
dependence corresponding to one gluon exchange. With present data, however, we
have no evidence for the additional factor $1/\log(\Lambda r)$, which according
to QCD should be introduced by the running of the coupling constant. The
expected part of the potential linear in $r$ is not seen. Probably the
bottomonia are too small to reach sufficiently far into the asymptotic region
of linear confinement. Perhaps a more flexible potential would exhibit the
linear part.

\end{document}